\input{aipcheck}

\documentclass[
  ,numberedheadings 
  ]
  {aipproc}

\layoutstyle{6x9}

\usepackage{psfrag}

\newcommand{\Pslash}{P\hspace{-0.25cm}/ \hspace{0.05cm}}

\begin{document}

\begin{flushright}
MPP-2009-173
\end{flushright}
\vspace{-0.1cm}

\title{Thermal masses in leptogenesis}


\author{Clemens P.~Kie\ss ig}{
  address={Max-Planck-Institut f\"{u}r Physik, F\"{o}hringer Ring 6,
  80805 M\"{u}nchen, Germany} }

\author{Michael Pl\"{u}macher}{
  address={Max-Planck-Institut f\"{u}r Physik, F\"{o}hringer Ring 6,
80805 M\"{u}nchen, Germany}
}


\begin{abstract}
We investigate the validity of using thermal masses in the kinematics
of final states in the decay rate of heavy neutrinos in leptogenesis
calculations. We find that using thermal masses this way is a
reasonable approximation, but corrections arise through quantum
statistical distribution functions and leptonic quasiparticles.



\end{abstract}

\maketitle


\section{Introduction}

Leptogenesis \cite{Sakharov,FukugitaYanagida86} is an extremely
successful theory in explaining the baryon asymmetry of the universe
by adding three heavy right-handed neutrinos $N_i$ to the standard model,
\begin{equation}
\delta {\mathcal L} = \bar{N}_i i \partial_\mu \gamma^\mu N_i - \lambda_\nu^{i
  \alpha} \bar{N}_i \phi^\dagger \ell_\alpha - \frac{1}{2} M_i
\bar{N}_i N_i^c + h.c.,
\end{equation}
with masses $M_i$ at the GUT-scale and Yukawa-couplings
$\lambda_\nu^{i \alpha}$ similar to the other fermions. This also
solves the problem of the light neutrino masses
via the see-saw mechanism without fine-tuning.

The heavy neutrinos decay into lepton and Higgs after inflation, the
decay is out of equilibrium since there are no gauge couplings to the
SM. If the CP asymmetry in the Yukawa-couplings is large enough, a
lepton asymmetry is created by the decays which is then partially
converted into a baryon asymmetry by sphaleron processes. As
temperatures 
are extremely
high, interaction rates need to be calculated using thermal field
theory (TFT) \cite{Giudice} rather than vacuum quantum field theory.





\section{Thermal corrections}

\subsection{Thermal masses}


When using bare thermal propagators in TFT
\cite{Bellac}, one can encounter IR singularities and gauge dependent
results. In order to cure this problem, the hard thermal loop (HTL)
resummation technique has been invented
\cite{BraatenPisarski90a,BraatenPisarski90b},
where for soft momenta $K \ll T$, resummed 
propagators are used. For a scalar
field,
\begin{equation}
i \Delta^* = i \Delta + i \Delta (-i \Sigma) i \Delta + ... =
\frac{i}{\Delta^{-1}-\Sigma}= \frac{i}{K^2 - m^2 - \Sigma}.
\end{equation}
The self energy $\Sigma$ then acts as a thermal mass $m_{\sf
th}^2(T):= m^2 + \Sigma$.

\subsection{May you put in thermal masses ``by hand''?}

Consider 
the decay rate $\Gamma(N_1 \rightarrow L H)$. The decay density which
appears in the Boltzmann equations is
\begin{equation}
\gamma_D=\int {\rm d} \tilde{p}_N {\rm d} \tilde{p}_L {\rm d} \tilde{p}_H
(2 \pi)^4 \delta^4 (P_N-P_L-P_H) | \mathcal{M} |^2 f_N (1+f_H) (1-f_L),
\end{equation}
where 
$
{\sf d} \tilde{p}_i = \frac{p_i^3}{(2 \pi)^3 2 E_i}
$
and we are using Fermi-Dirac and Bose-Einstein statistics with
Fermi blocking $(1-f)$ and Bose enhancement factors $(1+f)$. Thermal
masses are put in ``by hand'' and appear in the disperson relation of
Higgs and lepton doublet 
$
E_{H,L}=\sqrt{m_{H,L}^2(T) + p_{H,L}^2}.
$
The decay is only allowed if $M_N>m_L(T)+m_H(T)$.

Ref.~\cite{Giudice} included thermal masses this way, but chose
to approximate the quantum statistical distribution functions with
Maxwell-Boltzmann distributions, where
blocking and enhancement factors disappear. The deviation of
interaction rates from the quantum statistical case is typically
around 20--30\%.
This ad-hoc treatment of thermal masses seems intuitive and
plausible, but requires a closer look since
TFT describes Green's functions rather than external states and their
kinematics.

\begin{figure}
\label{optical}
\vspace{-0cm}
\psfrag{Right}{$\rightarrow$}
\scalebox{0.55}{\includegraphics{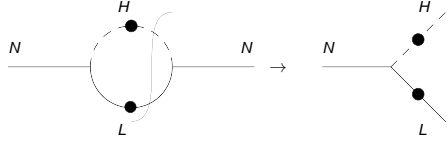}}
\caption{Neutrino decay rate via the optical theorem}
\end{figure}

The consistent way \cite{Bellac,Weldon83,KobesSemenoff86} to deal with
TFT effects on external states is by looking at the discontinuities of
the corresponding loop diagrams, where all propagators are treated
thermally and the external states like vacuum states. 
We consider the self-energy of the heavy neutrino (Fig.~\ref{optical})
and resum the Higgs and the lepton propagator using the HTL
resummation technique \cite{Bellac} in order to account for thermal
masses. In the imaginary time formalism, the self-energy is
\begin{equation}
\Sigma(N_1) = - (\lambda_\nu^\dagger \lambda_\nu)_{11} T \displaystyle\sum_{p_L^0=2 \pi {\sf i} n T} \int 
\frac{{\sf d}^3
p_L}{(2 \pi)^3} S^*(P_L) D^*(P_H).
\end{equation}
The resummed Higgs propagator is
$
D^*(P_H)=\frac{1}{P_H^2-m_H^2(T)},
$
where\\
$
{m_H^2(T)=(\frac{3}{16} g_2^2+ \frac{1}{16} g_Y^2 +
\frac{1}{4} y_t^2 + \frac{1}{2} \lambda) T^2}.
$
The resummed lepton propagator is
\begin{equation}
\label{leptonprop}
S^*(K)=\frac{1}{2 D_+(K)} (\gamma_0-\hat{\bf k} \cdot {\bf \gamma})
+\frac{1}{2 D_-(K)} (\gamma_0+\hat{\bf k} \cdot {\bf \gamma}),
\end{equation}
where
\begin{equation}
D_\pm(K)=-k_0 \pm  k + \frac{m_L(T)^2}{k} \left ( \pm1 - 
\frac{\pm k_0 - k}{2k} \ln \frac{k_0+k}{k_0-k}  \right )
\end{equation}
and 
$
m_L^2(T)=(\frac{3}{32} g_2^2+ \frac{1}{32} g_Y^2) T^2.
$
For simplicity, we approximate the lepton propagator with
$
{S^*(k)=\frac{{\not k}}{k^2-m_L^2(T)}}
$ \cite{Bellac}
and come back to the full HTL-result later.

According to finite-temperature cutting rules
\cite{Bellac,Weldon83,KobesSemenoff86}, the discontinuity of the self-energy is
\begin{equation}
{\sf tr}({\Pslash}_N \, {\sf Im} \, \Sigma) = - \int {\sf d} \tilde{p}_L
{\sf d} \tilde{p}_H (2 \pi)^4 \delta^4(P_N-P_H(T)-P_L(T)) \times
|\mathcal{M}|^2 (1-f_L+f_H),
\end{equation}
where the factor $1-f_L+f_H=(1-f_L)(1+f_H)+f_L f_H$ accounts for both
decays $N \rightarrow LH$ and inverse decays $LH \rightarrow N$. The
interaction rate
$
\Gamma=-\frac{1}{2 E_N} \, {\rm tr}(\Pslash_N \, {\rm Im} \, \Sigma)
$
gives the rates for decays and inverse decays,
$
\Gamma_D=(1-f_N^{\rm eq}(E_N)) \Gamma
$
and
$
\Gamma_{ID}=f_N^{\rm eq}(E_N) \Gamma,
$
where $f_N^{\rm eq}$ denotes the equilibrium distribution function.
The (inverse) decay density in the Boltzmann equation is
\begin{equation}
\gamma_{D,ID} = \int \frac{{\rm d^3} p_N}{(2 \pi)^3} f_N(E_N) \Gamma_{D,ID}\, ,
\end{equation}
where $f_N$ needs to be out of equilibrium \cite{Sakharov}.

The result equals the intuitive ``per hand'' treatment with two
caveats: The correct quantum statistical distribution functions always appear
in TFT even without including thermal masses, therefore it seems
inconsistent to use thermal masses and Maxwell-Boltzmann approximation
at the same time, as has been done by ref.~\cite{Giudice}. Moreover, the
HTL lepton propagator (eq.~\ref{leptonprop}) has a different structure
than the approximate propagator.

\subsection{Two leptonic quasiparticles}

\begin{figure}
\label{omegapm}
\caption{Two lepton dispersion relations, the usual dispersion relation in 
red.}
\psfrag{op}{\normalsize $\omega_+$}
\psfrag{om}{\normalsize $\omega_-$}
\psfrag{sqrtofp2m}{\normalsize $\sqrt{k^2+m_{\rm L}^2}$}
\scalebox{0.7}{\includegraphics{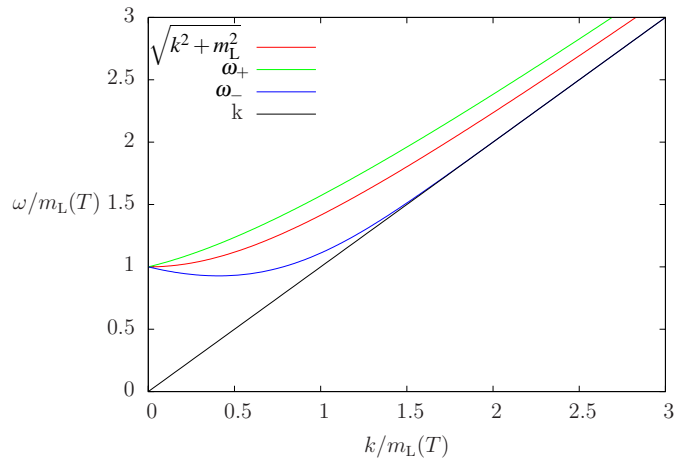}}
\end{figure}

The full HTL resummed lepton propagator (eq.~\ref{leptonprop}) has two
poles at $D_\pm(\omega_\pm,k)=0$, which correspond to two different
dispersion relations $\omega_\pm(k)$ (Fig.~\ref{omegapm})
\mbox{\cite{Klimov:1981ka,Weldon:1982aq,Weldon:1982bn,Weldon:1989ys}}.
For very large momenta, $\omega_+ \rightarrow \sqrt{k^2+ 2\, m_L^2(T)}$
and $\omega_- \rightarrow k$. In this limit, $\omega_+$
corresponds to a thermal mass of $\sqrt{2}\, m_L(T)$ and $\omega_-$ to a
massless mode. The decay rate for the heavy mode $\omega_+$ will thus be
reduced and the threshold for this decay will be at
${M_N=m_H(T)+\sqrt{2}\, m_L(T)}$, whereas the decay in the light mode
will be possible up to $M_N=m_H(T)$. The detailed shape of the decay
rate including the two leptonic modes and its influence on the
Boltzmann equations will be examined in a later work
\cite{KiessigPlumacher09}.

\section{Conclusions}

Using HTL resummation and the optical theorem, we confirm that putting
in thermal masses ``by hand'' as has been done by ref.~\cite{Giudice} is a
reasonable approximation. However, since quantum statistical
distribution functions always appear in the TFT treatment, it is
inconsistent to use thermal masses and Maxwell-Boltzmann distributions
at the same time. Moreover, the full HTL lepton propagator gives two
leptonic quasiparticles and will have an effect on the decay rate
which has to be examined in a future work \cite{KiessigPlumacher09}.

Looking further ahead it also seems desirable to look at thermal corrections
to scattering processes and the CP asymmetry in a consistent thermal
field theoretic treatment.


\begin{theacknowledgments}

We would like to thank Georg Raffelt, Markus Thoma, Dietrich
B\"{o}deker, Florian Hahn-W\"{o}rnle, Steve Blanchet and Denis Besak for
fruitful and inspiring discussions.

\end{theacknowledgments}



\bibliographystyle{aipproc}   

\bibliography{paper}

\IfFileExists{\jobname.bbl}{}
 {\typeout{}
  \typeout{******************************************}
  \typeout{** Please run "bibtex \jobname" to optain}
  \typeout{** the bibliography and then re-run LaTeX}
  \typeout{** twice to fix the references!}
  \typeout{******************************************}
  \typeout{}
 }

\end{document}